\documentclass[aps,prl,twocolumn,groupedaddress,showpacs]{revtex4}

\usepackage{graphics}
\usepackage{amssymb,amsmath}
\newcommand{\beq}{\begin{equation}}
\newcommand{\eeq}{\end{equation}}
\newcommand{\bea}{\begin{eqnarray}}
\newcommand{\eea}{\end{eqnarray}}
\newcommand{\non}{\nonumber}
\newcommand{\mbf}{\mathbf}
\newcommand{\trm}{\textrm}

\newcommand{\Q}{\mbf{Q}}
\newcommand{\n}{\mbf{n}}
\newcommand{\x}{\mbf{x}}
\newcommand{\z}{\mbf{z}}

\newcommand{\CL}[1][L]{\left<C_{#1}\right>_N}

\newcommand{\QNL}{{\cal Q}_L^N}
\newcommand{\PNL}{{\cal P}_L^N}
\newcommand{\PL}{(P_L)_{l_1l_2}^j}
\renewcommand{\P}{(P_L)}

\newcommand{\AL}{(A_L^j)_{l_1l_2}}
\newcommand{\BL}{(B_L^h)_{k_1k_2}}
\newcommand{\PpL}{(P'_L)_{k_1k_2}^h}

\newcommand{\refrule}  {({\sc i})}
\newcommand{\refinput} {({\sc ii})}
\newcommand{\refstable}{({\sc iii})}
\newcommand{\refinvert}{({\sc iv})}

\begin{document}

\preprint{LU TP 02-38}

\title{The Number of Attractors in Kauffman Networks}


\author{Bj\"orn Samuelsson}
\email[]{bjorn@thep.lu.se}
\author{Carl Troein}
\email[]{carl@thep.lu.se}

\affiliation{
Complex Systems Division, Department of Theoretical Physics\\
Lund University,  S\"olvegatan 14A,  S-223 62 Lund, Sweden}


\date{January 17, 2003}

\begin{abstract}
The Kauffman model describes a particularly simple class of random Boolean
networks. Despite the simplicity of the model, it exhibits complex behavior
and has been suggested as a model for real world network problems. We
introduce a novel approach to analyzing attractors in random Boolean
networks, and applying it to Kauffman networks we prove that the average
number of attractors grows faster than any power law with system size.
\end{abstract}


\maketitle


\section{Introduction}
   
We are increasingly often faced with the problem of modeling complex
systems of interacting entities, such as social and economic networks, 
computers on the Internet, or protein interactions in living cells.
Some properties of such systems can be modeled by Boolean networks.
The appeal of these networks lies in the finite (and small) number of
states each node can be in, and the ease with which we can handle
the networks in a computer.

A deterministic Boolean network has a finite number of states. Each
state maps to one state, possibly itself. Thus, every network has at
least one cycle or fixed point, and every trajectory will lead to such
an attractor.  The behavior of attractors in Boolean networks has been
investigated extensively, see e.g. \cite{Bornholdt:98, Lemke:01,
Bhattacharjya:96,Bagley:96, Fox:01, Oosawa:02}. For a recent
review, see \cite{Aldana:02}.

A general problem when dealing with a system is finding the set of
attractors. For Boolean networks with more than a handful of nodes,
state space is too vast to be searched exhaustively. In some
cases, a majority of the attractor basins are small and very hard
to find by random sampling. One such case is the Kauffman model
\cite{Kauffman:69}.
Based on experience with random samplings, it has been commonly believed
that the number of attractors in that model grows like the square root of
system size. Lately this has been brought into question
\cite{Bastolla:97,Bastolla:98,Bilke:01,Socolar:02}.
Using an analytic approach, we are able to prove that the number
attractors grows faster than any power law.
The approach is based on general probabilistic reasoning that may
also be applied to other systems than Boolean networks.
The derivations in the analysis section are just one application of
the approach. We have attempted to focus on the method and our
results, while keeping the mathematical details at a minimum level
needed for reproducibility using standard mathematical methods.

In 1969 Kauffman introduced a type of Boolean networks as a model for gene
regulation \cite{Kauffman:69}. These networks are known as $N$-$K$
models, since each of the $N$ nodes has a fixed number of inputs $K$.
A Kauffman network is synchronously updated, and the state (0 or 1) of
any node at time step $t$ is some function of the state of its input nodes
at the previous time step. An assignment of states to all nodes is referred
to as a {\it configuration}.
When a single network, a {\it realization}, is created, the choice of input
nodes and update functions is random, although the update functions are
not necessarily drawn from a flat distribution. This reflects a null
hypothesis as good or bad as any, if we have no prior knowledge of the
details of the networks we wish to model.

In this letter, we will first present our approach,
and then apply it to Kauffman's
original model, in which there are 2 inputs per node and the same probability
for all of the 16 possible update rules. These 16 rules are the Boolean
operators of two or fewer variables: {\sc and}, {\sc or}, {\sc true}, etc.
This particular $N$-$K$ model falls on the {\it critical line}, where
the network dynamics is neither ordered nor chaotic \cite{Derrida:86a,
Derrida:86b, Bastolla:97}.

\section{Approach}

Our basic idea is to focus on the problem of finding the number of cycles of
a given length $L$ in networks of size $N$. As we will see, the discreteness
of time makes it convenient to handle cycles as higher-dimensional fixed
point problems. Then it is possible to do the probabilistic averaging in a way
which is suitable for an analytic treatment. This idea may also be expanded
to applications with continuous time, rendering more complicated but also more
powerful methods.

We use four key assumptions: \refrule{} the rules are chosen
independently of each other and of $N$, \refinput{} the input nodes
are independently and uniformly chosen from all $N$ nodes,
\refstable{} the dynamics is dominated by stable nodes, and
\refinvert{} the distribution of rules is invariant due to
inversion of any set of inputs. \refinvert{} means e.g. that the
fraction of {\sc and} and {\sc nor} gates are the same whereas the
fraction {\sc and} and {\sc nand} gates may differ. \refinvert{} is
presumably not necessary, but simplifies the calculations
drastically. \refstable{} is expected to be valid for any non-chaotic
network obeying \refrule{} and \refinput{} \cite{Flyvbjerg:88}. Note
that \refrule{} does not mean that the number of inputs must be the
same for every rule. We could write a general treatment of all models
obeying \refrule{} -- \refinvert{}, but for simplicity we focus on the
Kauffman model.

We will henceforth use $\CL$ to denote the expectation value
of the number of $L$-cycles over all networks of size $N$, with $L=1$
referring to fixed points.
The average number of fixed points, $\CL[1]$, is particularly simple to
calculate. For a random choice of rules, \refrule{} and \refinvert{}
imply that the output state of the net is independent of the input
state. Hence, the input and output states will on average coincide
once on enumeration of all input states. This means that $\CL[1]=1$.

The problem of finding other $L$-cycles can be transformed to a fixed
point problem. Assume that a Boolean network performs an $L$-cycle.
Then each node performs one of $2^L$ possible time series of output
values. Consider what a rule does when it is subjected to such time
series on the inputs. It performs some boolean operation, but it also
delays the output, giving a one-step difference in phase for the
output time series. If we view each time series as a state, we have a
fixed point problem. $L\CL$ is then the average number of input states
(time series), for the whole network, such that the output is the same
as the input.

To take advantage of assumption \refinvert{}, we introduce the notion of
$L$-cycle patterns. An $L$-cycle pattern is $s$ and $s$ inverted,
where $s$ is a time series with period $L$. Let $\Q$ denote a choice
of $L$-cycle patterns for the net, and let $P(\Q)$ denote the
probability that the output of the net is $\Q$. Using the same line of
reasoning as for fixed points, we conclude that \refrule{} and \refinvert{}
yield
\beq
  \CL = \frac1{L}\sum_{\Q \in \QNL} P\left(\Q\right)
  \label{eq: CLN(P(Q))}
\eeq
where $\QNL$ is the set of proper $L$-cycles of an $N$-node net.
A proper $L$-cycle has no period shorter than $L$.

\section{Analytic calculations}

Assumption \refinput{} implies that $P(\Q)$ is invariant under permutations
of the nodes. Let $\n=(n_0,\ldots,n_{m-1})$ denote the number of nodes
expressing each of the $m=2^{L-1}$ patterns. For $n_j$, we refer to $j$
as the pattern index. For convenience, let the constant pattern have index
$0$. Then
\beq
  \CL = \frac1L\sum_{\n \in \PNL} 
           \binom{N}{\n} P\left(\Q\right)
  \label{eq: CL multinom}
\eeq
where $\binom{N}{\n}$ denotes the multinomial $N!/(n_0!\cdots
n_{m-1}!)$ and $\PNL$ is the set of partitions $\n$ of $N$ such that
$\Q \in \QNL$. That is, $\n$ represents a proper $L$-cycle.

What we have this far is merely a division of the probability of an
$L$-cycle into probabilities of different flavors of $L$-cycles.
Now we assume that each node has $2$ inputs. Then, we get a simple
expression for $P(\Q)$ that inserted into Eq. \eqref{eq: CL multinom}
yields
\bea
  \CL &=&\frac{1}{L}\sum_{\n \in \PNL} 
           \binom{N}{\n}\non\\
         &&\times \prod_{\substack{ 0 \leq j < m\\
                                 n_j \neq 0}}
          \left(\sum_{0\leq l_1,l_2<m}\frac{n_{l_1}n_{l_2}}{N^2}\PL\right)
         ^{\textstyle n_j}
  \label{eq: CL-exact}
\eea
where $\PL$ denotes the probability that the output pattern of a
random 2-input rule has index $j$, given that the input patterns have
the indices $l_1$ and ${l_2}$ respectively. Note that Eq. \eqref{eq:
CL-exact} is an exact expression for the average number of proper
$L$-cycles in an $N$-node random Boolean network which satisfies the
assumptions \refrule{}, \refinput{}, \refinvert{}, and where each node
has $2$ inputs.

From now on, we only consider the Kauffman model, meaning that we also
restrict the distribution of rules to be uniform. It is
instructive to explore some properties of $\PL$; these will also be
needed in the following calculations.  We see that
\beq
  \P_{00}^0 = 1,~~~\P_{l_10}^0=\tfrac12~~~\trm{and}~~~
        \P_{l_1l_2}^0 \geq \tfrac18
  \label{eq: Pt0}
\eeq
for $1\leq l_1,l_2<m$. Further, we note that for a given $j\neq0$,
$\P_{l_10}^j$ has a non-zero value for exactly one $l_1 \in
\{1,\ldots,m-1\}$. Let $\phi_L(j)$ denote that value of $l_1$.
We can see $\phi_L$ as a function that rotates an $L$-cycle pattern
one step backwards in time. With this in mind we define $\phi_L(0)=0$.
Now, we can write
\beq
  \P_{l_10}^j = \tfrac12\delta_{l_1\phi_L(j)}
  \label{eq: Ptl0}
\eeq
for $1\leq j<m$. ($\delta$ is the Kronecker delta.)

We can view $\phi_L$ as a permutation on the set $\{0,\ldots,m-1\}$.
Thus, we divide this index space into permutation cycles which are
sets of the type $\{j,\phi_L(j),\phi_L\circ\phi_L(j),\ldots\}$.  We
refer to these permutation cycles as invariant sets of $L$-cycles.
Let $\rho_L^0,\ldots,\rho_L^{H_L-1}$ denote the invariant sets of
$L$-cycles, where $H_L$ is the number of such sets. For convenience,
let $\rho_0$ be the invariant set $\{0\}$. If two $L$-cycle patterns
belong to the same invariant set, they can be seen as the same
pattern except for a difference in phase.

We want to find the behavior of $\CL$, for large $N$, by approximating
Eq. \eqref{eq: CL-exact} with an integral. To do this, we use Stirling's
formula $n! \approx (n/e)^n\sqrt{2\pi n}$ while noting that the
boundary points where $n_j=0$ for some $j$ can be ignored in the integral
approximation. Let $x_j = n_j/N$ for $j=0,\ldots,m-1$ and
integrate over $\x = (x_1,\ldots,x_{m-1})$. $x_0$ is implicitly set to
$x_0 = 1 - \sum_{j=1}^{m-1} x_j$. We get
\beq
  \CL \approx \frac1L\left(\frac{N}{2\pi}\right)^{(m-1)/2}
     \mspace{-30mu}\int\displaylimits_{0 < x_0,\ldots,x_{m-1}}
     \mspace{-25mu}d\x\mspace{5mu}
     \frac{e^{Nf_L(\x)}}{\prod_{j=0}^{m-1}\sqrt{x_j}}
  \label{eq: CL approx}
\eeq
where
\beq
  f_L(\x) = \sum_{j=0}^{m-1}
    x_j \ln\left(\frac{1}{x_j}
    \sum_{0\leq l_1,l_2<m}x_{l_1}x_{l_2}\PL\right)~.
  \label{eq: fL}
\eeq
Eq. \eqref{eq: fL} can be seen as an average $\left<lnX\right>$, where
$X$ is the expression inside the parentheses. Hence, the concavity
of $x \rightarrow \ln x$ gives
$f_L(\x) = \left<lnX\right> \leq ln\left<X\right> = 0$ with equality if
and only if
\beq
  x_j = \sum_{0\leq l_1,l_2<m}x_{l_1}x_{l_2}\PL
  \label{eq: f-zeros}
\eeq
for all $j=0,\ldots,m-1$.

Note that Eq. \eqref{eq: f-zeros} can be interpreted as a mean-field
equation of the model. Using Eq. \eqref{eq: f-zeros} for $j=0$ and
Eq. \eqref{eq: Pt0} we see that $f_L(\x)$ comes arbitrarily close to
zero only in the vicinity of $\x=0$, and for large $N$, the relevant
contributions to the integral in Eq. \eqref{eq: CL approx} come from
this region. Thus, the dynamics of the net is dominated by stable
nodes, in agreement with \cite{Flyvbjerg:88, Bilke:01}. This
means that assumption \refstable{} is satisfied by the Kauffman model. Using
Eqs. \eqref{eq: Pt0} and \eqref{eq: Ptl0}, a Taylor-expansion of
$f_L(\epsilon\x)$ yields
\bea
  f_L(\epsilon\x) &=& \epsilon\sum_{j=1}^{m-1}x_j\ln\frac{x_{\phi(j)}}{x_j}
                + \epsilon^2\sum_{j=0}^{m-1}x_j\frac{\x\cdot A_L^j\x}{x_{\phi(j)}} \non\\
                &&- \frac{\epsilon^3}2\sum_{j=1}^{m-1}x_j
                 \left(\frac{\x\cdot A_L^j\x}{x_{\phi(j)}}\right)^2
                +O\left(\epsilon^4\right)
  \label{eq: Taylor-x}
\eea
where $\AL = \PL - \frac12(\delta_{l_1\phi(j)} + \delta_{l_2\phi(j)})$.

The first order term of Eq. \eqref{eq: Taylor-x} has $0$ as its maximum
and reaches this value if and only if $x_{\phi_L(j)}=x_j$ for all
$j=1,\ldots,m-1$. The second order term is zero at these points, while
the third order term is less than zero for all $\x\neq0$. Hence, the
first and third order terms are governing the behavior for large $N$.
Using the saddle-point approximation, we reduce the integration space
to the space where the first and second order terms are $0$. Let
$z_h=N^{1/3}\sum_{j\in{\rho_L^h}}x_j$ for $h=1,\ldots,H_L-1$ and
let $\PpL$ denote the probability that the output pattern of a random
rule belongs to $\rho_L^h$, given that the input patterns are randomly
chosen from $\rho_L^{k_1}$ and $\rho_L^{k_2}$ respectively.
Thus, we approximate Eq. \eqref{eq: CL approx} for large $N$ as
\beq
  \CL \approx \alpha_L\beta_LN^{\gamma_L}
  \label{eq: asymptote}
\eeq
where
\bea
  \alpha_L &=& \left(L\prod_{h=1}^{H_L-1}\left|\rho_L^h\right|\right)^{-1}
                \left(\frac1{2\pi}\right)^{(H_L-1)/2}
\\
  \beta_L &=& \mspace{-30mu}\int\displaylimits_{\mspace{7mu}0 < z_1,\ldots,z_{H_L-1}\mspace{-7mu}}
             \mspace{-35mu}d\z\mspace{5mu}
	     \frac {\displaystyle \exp\left(\displaystyle 
                      -\frac12\sum_{h=1}^{H_L-1}\frac1{z_h}
                    \left(\z\cdot B_L^h\mbf{z}\right)^2\right)}
             {\displaystyle \prod_{h=1}^{H_L-1}\sqrt{z_h}}
  \label{eq: beta-integral}
\eea
\bea
  \gamma_L &=& \frac{H_L-1}3
  \label{eq: gamma_L}
\eea
and $\BL = \PpL-\frac12\left(\delta_{h_1h}+\delta_{h_2h}\right)$.
($|\rho|$ denotes the number of elements of the set $\rho$.)

$H_L$ grows rapidly with $L$. The number of elements in an invariant
set of $L$-cycle patterns is a divisor of $L$. If an invariant set
consists of only one pattern, it is either the constant pattern,
or the pattern with alternating zeros and ones. The latter
is only possible if $L$ is even. Thus, $H_L - 1 \geq
(2^{L-1}-1)/L$, with equality if $L$ is a prime number $>2$. Applying
this conclusion to Eqs. \eqref{eq: gamma_L} and \eqref{eq: asymptote}, we
see that for any power law $N^\xi$, we can choose an $L$ such that
$\CL$ grows faster than $N^\xi$.

\section{Numerical results}

\begin{figure}[b]
\includegraphics{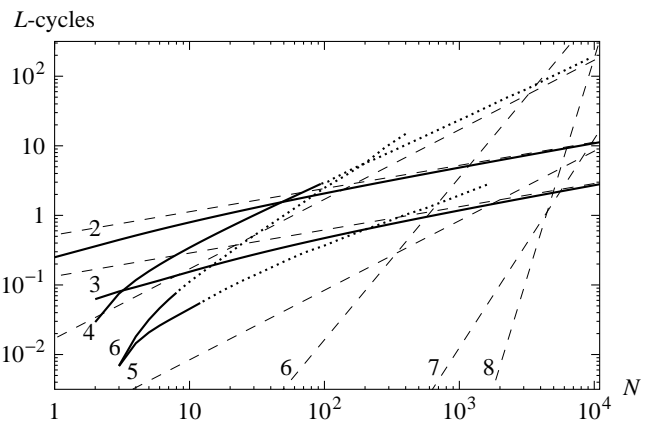}
\caption{\label{fig:cycles}
The number of $L$-cycles as functions of the network size for
$2 \le L \le 6$. The numbers in the figure indicate $L$.
Dotted lines are used for values obtained by Monte Carlo
summation, with errors comparable to the line width.
The asymptotes for $N \le 8$ have been included as dashed lines. Their
slopes are $\gamma_2=\gamma_3=\frac13$, $\gamma_4=\gamma_5=1$,
$\gamma_6=\frac73$, $\gamma_7=3$, and $\gamma_8=\frac{19}{3}$. }
\end{figure}

We have written a set of programs to compute the number of $L$-cycles
in Kauffman networks, Eq. \eqref{eq: CL-exact},
both by complete enumeration and using Monte Carlo methods, and tested
them against complete enumeration of the networks with
$N \le 4$.
The results for $2 \le L \le 6$ are shown in Fig. \ref{fig:cycles}, along
with the corresponding asymptotes. The asymptotes were obtained by
Monte Carlo integration of Eq. \eqref{eq: beta-integral}. For low $N$,
$\CL$ is dominated by the boundary points neglected in
Eq. \eqref{eq: CL approx}, and its qualitative behavior is not obvious in
that region.

A straighforward way to count attractors in a network is to simulate
trajectories from random configurations and count distinct target
attractors. As has been pointed out in \cite{Bilke:01}, this gives a biased
estimate. Simulations in \cite{Kauffman:69} with up to 200 trajectories
per network indicated $\sqrt{N}$ scaling with system size, whereas
\cite{Bilke:01} reported linear behavior with 1000 trajectories.
That the true growth is faster than linear has now been firmly
established \cite{Socolar:02}.

\begin{figure}[bt]
\includegraphics{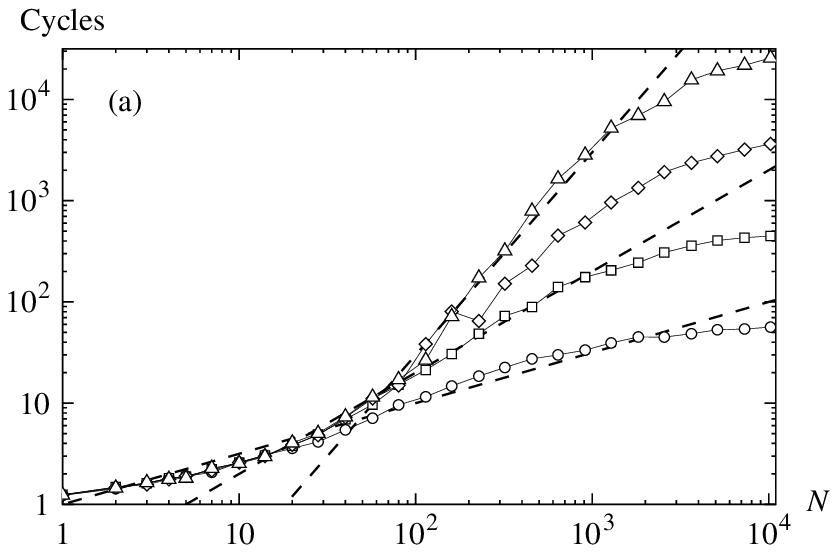}
\includegraphics{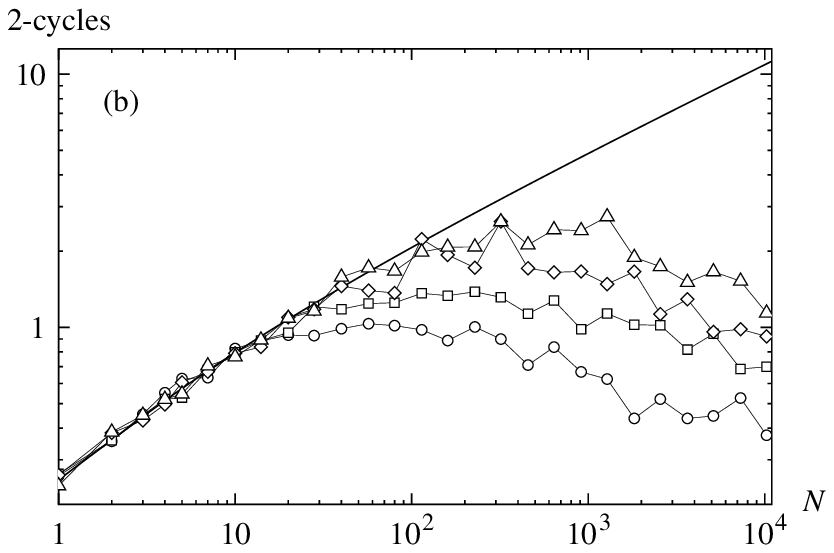}
\caption{\label{fig:sim}
The number of observed attractors (a) and 2-cycles (b) per network
as functions of $N$ for different numbers of trajectories $\tau$:
100(circles), $10^3$(squares), $10^4$(diamonds), and $10^5$(triangles).
In (a), the dashed lines have slopes 0.5, 1, and 2.
In (b) the solid line shows $\CL[2]$, the true number of 2-cycles.
In the high end of (b), $1 \sigma$ corresponds to roughly $20\%$.
}
\end{figure}

To closer examine the problem of biased undersampling, we have
implemented the network reduction algorithm from \cite{Bilke:01},
and gathered statistics on networks with $N \lesssim 10^4$.
We repeated the simulations for different numbers of trajectories $\tau$,
with $100\le\tau\le10^5$.
For each $N$ and $\tau$, $10^3$ network realizations were examined,
and we discarded configurations if no cycle was found within $2^{13}$
time steps. The results are summarized in Fig. \ref{fig:sim}a.

For $\tau=100$, the number of attractors follows $\sqrt{N}$ remarkably well,
considering that $\tau=10^3$ gives the quite different $N$ behavior seen
in \cite{Bilke:01}.
If we extrapolate wildly from a log--loglog plot, $e^{0.3\sqrt{N}}$ fits
the data rather well.

As another example of how severe the biased undersampling is,
we have included a plot of the number of 2-cycles found in the
simulations (Fig. \ref{fig:sim}b). The underlying distribution is less
uniform than for the total number of attractors, so the errors are
larger, but not large enough to obscure the qualitative behavior.
The number of observed 2-cycles is close to $\CL[2]$ for low $N$, but as
$N$ grows, a vast majority of them are overlooked. As expected, this
problem sets in sooner for lower $\tau$, although the difference is not
as marked as it is for the total number of attractors.


\section{Summary}
We have introduced a novel approach to analyzing attractors of
random Boolean networks. By applying it to Kauffman networks,
we have proven that the number of attractors in these grows
faster than any power law with network size $N$. This is in sharp
contrast with the previously cited $\sqrt{N}$ behavior, but in
agreement with recent findings.

For the Kauffman model, we have derived an expression for the
asymptotic growth of the number of $L$-cycles, $\CL$. This
expression is corroborated by statistics from network simulations.
The simulations also demonstrate that biased undersampling of state
space is a good explanation for the previously observed $\sqrt{N}$ behavior.

\begin{acknowledgments}
We wish to thank Carsten Peterson, Bo S\"oder\-berg, and Patrik Ed\'en
for valuable discussions.
This work was in part supported by the National Research School in
Genomics and Bioinformatics.
\end{acknowledgments}


\end{document}